\documentclass[floats,floatfix,amssymb,prl,twocolumn,superscriptaddress,nofootinbib]{revtex4-1}
\usepackage{amssymb,amsmath,verbatim,mathtools,needspace,enumitem,etoolbox,graphicx,physics,microtype,afterpage,bm}
\usepackage[dvipsnames, usenames]{xcolor}
\definecolor{linkcolor}{rgb}{0.0,0.3,0.5}
\usepackage[unicode, colorlinks=true, linkcolor=linkcolor, citecolor=linkcolor, filecolor=linkcolor,urlcolor=linkcolor, pdfusetitle]{hyperref}
\usepackage[all]{hypcap}
\usepackage[T1]{fontenc}
\usepackage[utf8]{inputenc}
\usepackage{tabularx}
\usepackage{float}
\def\be{\begin{equation}}
\def\ee{\end{equation}}
\def\bea{\begin{eqnarray}}
\def\eea{\end{eqnarray}}
\def\Aa{\frac{a^{\prime}}{a}}
\def\Ab{\Big(\frac{a^{\prime}}{a}\Big)^2}
\def\Ac{\frac{a^{\prime\prime}}{a}}

\def\La{\partial_i \,\partial^i}

\def\deu{{\delta}_1}
\def\ded{{\delta}_2}

\def\beq{\begin{equation}}
\def\eeq{\end{equation}}
\def\d{{\rm d}}
\def\PBH{\text{\tiny PBH}}

 at 10truept

\newcommand{\llp}{\left [}
\newcommand{\rrp}{\right ]}
\newcommand{\lp}{\left (}
\newcommand{\rp}{\right )}

 at 10truept

\usepackage{lmodern}
\allowdisplaybreaks
\usepackage{tikz}
\usepackage{color}
\usepackage{framed}
\usepackage{hyperref}
\hypersetup{colorlinks, citecolor=verdes, linkcolor=black, urlcolor=verdes}
\definecolor{rossos}{cmyk}{0,1,1,0.55}
\definecolor{bluscuro}{rgb}{0.15, 0.2, .85}
\definecolor{bluchiaro}{cmyk}{1,.3,0.,0.1}
\definecolor{ForestGreen}{rgb}{0.13, 0.55, 0.13}
\definecolor{verdes}{rgb}{0.1, 0.5, 0.1}%
\renewcommand{\d}{{\rm d}}

\def\lsim{\mathrel{\rlap{\lower4pt\hbox{\hskip0.5pt$\sim$}}
    \raise1pt\hbox{$<$}}}         
\def\gsim{\mathrel{\rlap{\lower4pt\hbox{\hskip0.5pt$\sim$}}
    \raise1pt\hbox{$>$}}}         

\def\d{{\rm d}}

\def\PBH{\text{\tiny PBH}}

\def\com{\text{\tiny  com}}

\definecolor{myforestgreen}{rgb}{0.13, 0.55, 0.13}

\begin{document}

\title{How well do we know the primordial black hole abundance? \\The crucial role of nonlinearities when approaching the horizon}

\author{Valerio De Luca}
\email{vdeluca@sas.upenn.edu}
\affiliation{Center for Particle Cosmology, Department of Physics and Astronomy,
University of Pennsylvania 209 S. 33rd St., Philadelphia, PA 19104, USA}

\author{Alex Kehagias}
\email{kehagias@central.ntua.gr}
\affiliation{Physics Division, National Technical University of Athens, 15780 Zografou Campus, Athens, Greece}
\affiliation{CERN, Theoretical Physics Department, Geneva, Switzerland}

\author{Antonio Riotto}
\email{antonio.riotto@unige.ch}
\affiliation{D\'epartement de Physique Th\'eorique and Gravitational Wave Science Center (GWSC), Universit\'e de Gen\`eve, CH-1211 Geneva, Switzerland CH-1211 Geneva, Switzerland}

\begin{abstract}
\noindent
 We discuss  the  non-linear corrections entering in the calculation of the primordial black hole abundance from   the 
 non-linear radiation transfer function and 
the determination of the true physical horizon crossing. We show that the current standard techniques to calculate the abundance of primordial black holes suffer from   uncertainties and argue that the primordial black hole abundance may be much smaller than what routinely considered.
This would imply, among other consequences, that the interpretation of the recent pulsar timing arrays data from scalar-induced gravitational waves may not be ruled out because of an overproduction of primordial black holes.
\end{abstract}

\maketitle

\noindent{{\bf{\it Introduction.}}}
The various   detections of gravitational waves (GWs) 
originated from the mergers of  black hole binaries~\cite{LIGOScientific:2016aoc, LIGOScientific:2018mvr,LIGOScientific:2020ibl, LIGOScientific:2021djp} have resurrected  the interest in the 
physics of Primordial Black Holes (PBHs)~\cite{Sasaki:2018dmp, Carr:2020gox, Green:2020jor}. Some of the LIGO/Virgo/KAGRA data may be in fact  of primordial origin \cite{Bird:2016dcv,Sasaki:2016jop,Clesse:2016vqa,Ali-Haimoud:2017rtz,Hutsi:2020sol, DeLuca:2021wjr,Franciolini:2021tla,Franciolini:2022tfm} 
and  forthcoming  GW experiments might  shed light on the possible existence of PBHs \cite{Chen:2019irf,Pujolas:2021yaw,DeLuca:2021hde,Barsanti:2021ydd,Bavera:2021wmw}.

In the standard scenario, PBHs  are considered  to be  born in the radiation-dominated phase  by the collapse of large overdensities created during inflation on small scales~\cite{Sasaki:2018dmp}. Upon horizon re-entry, the very same sizeable  fluctuations generate  GWs at second-order  in perturbation theory (see Ref. \cite{Domenech:2021ztg} for a review).

 The recent pulsar timing arrays (PTA) data releases by the NANOGrav~\cite{NG15-SGWB,NG15-pulsars}, EPTA~\cite{EPTA2-SGWB,EPTA2-pulsars,EPTA2-SMBHB-NP}, PPTA~\cite{PPTA3-SGWB,PPTA3-pulsars,PPTA3-SMBHB} and CPTA~\cite{CPTA-SGWB} collaborations have shown  evidence for the presence of a stochastic background of GWs and have  raised the question if it can be ascribed to the scalar-induced GWs that may be sourced along with PBH formation. Whether or not this is possible depends crucially
on the PBH abundance. The argument goes as follows. The amount of GWs induced at second-order depends on the square of the amplitude of the dimensionless curvature perturbation power spectrum ${\cal P}_\zeta$, $\Omega_{\text{\tiny GW}}\sim {\cal P}_\zeta^2$. On the other hand, the abundance
of PBHs is exponentially  sensitive to the same amplitude,  $f_{\text{\tiny PBH}}\sim {\rm exp}(-1/{\cal P}_\zeta)$, where $f_{\text{\tiny PBH}}$ is the PBH abundance with respect to the total dark matter. A large stochastic background of GWs requires  large values of ${\cal P}_\zeta$,  automatically generating a PBH abundance  which tends to be too large to be compatible with the 15-year NANOGrav data, (unless some negative non-Gaussianity is introduced), see Refs.~\cite{Franciolini:2023pbf,Liu:2023ymk,Wang:2023ost,Cai:2023dls,Inomata:2023zup,Figueroa:2023zhu,Yi:2023mbm,Zhu:2023faa} for recent works along this direction.

Given what is at stake,  the compatibility of the first discovery of a stochastic background of GWs with the PBH scenario, but also  for more general reasons, a natural and fundamental question to ask is  how well do we know the PBH abundance to make any definite conclusion. 

The goal of this paper is to argue  that there are  sources of uncertainties  plaguing the standard calculation of the PBH abundance, mostly coming from the details of horizon crossing and the effect of the  non-linear  radiation  transfer function. These effects  might   change  the PBH abundance with respect to what  considered in the literature.

\vskip 0.5cm
\noindent
\noindent{{\bf{\it A quick summary of the  standard PBH abundance calculation.}}} In this section we briefly summarize what it is routinely done, and currently the best way to our knowledge,  to compute the PBH abundance in the literature, see for instance Refs.~\cite{Young:2019yug, Biagetti:2021eep}. 

As we already mentioned,  we will focus on the  PBH formation from the collapse of sizeable overdensities that are generated during inflation and re-enter the cosmological horizon during the subsequent radiation-dominated era. One key quantity is the  curvature perturbation $\zeta$ which appears in the metric in the comoving uniform-energy density gauge as
\be
{\rm d}s^2=-{\rm d}t^2+a^2(t)e^{2\zeta}{\rm d}{\bf x}^2,
\ee
where $a(t)$ is the scale factor in terms of the cosmic time.
On superhorizon scales, one applies the gradient expansion~\cite{Shibata:1999zs} to relate the non-linear density contrast $\delta^\com(r,t)$  on  comoving orthogonal slicings and the time independent curvature perturbation $\zeta(r)$ as~\cite{Harada:2015yda}
\be
\label{deltaNL}
\delta^\com(t,r)=-\frac{8}{9}\frac{1}{a^2H^2} e^{-5\zeta(r)/2}\nabla^2 e^{\zeta(r)/2},
\ee
where $H$ is the Hubble rate.
Whether or not  cosmological perturbations may  gravitationally collapse to  form a PBH depends on the amplitude measured at the peak of the compaction function, defined to be the mass excess compared to the background value in a given radius~\cite{Harada:2015yda, Musco:2018rwt,Escriva:2019phb}. On superhorizon scales it reads
\be
\mathcal{C} (r) = - \frac{2}{3}\, r\, \zeta' (r) \llp 2 +  r\, \zeta' (r) \rrp,
\ee
where the prime stands for differentiation with respect to $r$.
The compaction function has a maximum at the comoving length scale $r_m$ satisfying
\be
\mathcal{C}'(r_m) = 0 \qquad  {\rm or}  \qquad  \zeta' (r_m) + r_m \zeta'' (r_m) = 0.
\ee
One can therefore define  a smoothed perturbation amplitude as the volume average of the energy density contrast within the scale $r_m$ at the cosmological horizon crossing time $t_H$~\cite{Musco:2018rwt}
\be
\label{sh}
\delta_m=\frac{3}{\left(r_m e^{\zeta(r_m)}\right)^3}\int_0^{r_m} {\rm d} r\,\delta^\com(r,t_H)\left(r e^{\zeta(r)}\right)^2\left(r e^{\zeta(r)}\right)',
\ee
where a top-hat window function is adopted to account for the treatment of the threshold~\cite{Young:2019osy}.
This represents the main quantity determining the abundance of PBHs. When computed at the cosmological horizon crossing time $t_H$ (the perturbations, if large enough, collapse into a PBH very rapidly after horizon crossing)
\be
\epsilon(t_H)=\frac{r_H}{r_m e^{\zeta(r_m)}}=\frac{1}{r_m e^{\zeta(r_m)} aH }=1,
\ee
it  simplifies to
\be
\label{deltam}
\delta_m = \delta_l - \frac{3}{8} \delta_l^2, \qquad \delta_l = - \frac{4}{3}r_m \zeta' (r_m).
\ee
The PBH abundance is then  computed by integrating the probability distribution function of the smoothed density contrast from a threshold value $\delta_c$ on, as
\be
\beta = \int_{\delta_c} \lp \frac{M_\PBH}{M_H} \rp P(\delta_m) \d \delta_m,
\ee
in terms of the PBH mass $M_\PBH$ and the mass enclosed in the cosmological horizon at the horizon crossing time $M_H$.
However, from the relation shown above, one can use the conservation of probability to write 
\be
P(\delta_l) \d \delta_l = P(\delta_m) \d \delta_m,
\ee
such that the linear smoothed density contrast $\delta_l$ is the ultimate key parameter which we have to compute the probability of~\cite{Germani:2019zez,Young:2019yug,Biagetti:2021eep,DeLuca:2022rfz,Ferrante:2022mui,Gow:2022jfb}, with a corresponding threshold given by~\cite{Young:2019yug}
\be
\label{threhsold_delta}
\delta_{l,c} = \frac{4}{3} \lp 1- \sqrt{1-\frac{3}{2}\delta_c} \rp.
\ee
For Gaussian curvature perturbations, the probability for the linear density contrast is Gaussian and is exponentially sensitive to the threshold $\delta_{l,c}$ and the variance $\sigma^2_l$
\begin{align}
\label{sigma2}
P(\delta_l)&=\frac{\delta_{l,c}}{\sqrt{2\pi}\sigma_l} e^{-\delta_{l,c}^2/2\sigma_l^2},\nonumber\\
\sigma_l^2&= \int\frac{{\rm d}k}{k} {\cal P}_{\delta_l}(k)\nonumber\\
&=\frac{16}{81}\int\frac{{\rm d}k}{k}(k r_H)^4W^2(k,r_m) T^2(k r_H) {\cal P}_\zeta(k),
\end{align}
where $W(k,r_m$) is the Fourier transform of the top-hat window function in real space with radius $r_m$, and
\be
T(x)=3\frac{\sin(x/\sqrt{3})-(x/\sqrt{3})\cos(x/\sqrt{3})}{(x/\sqrt{3})^3}
\ee
is the linear radiation transfer function which has to be finally computed at the time of horizon crossing $t_H$.

\begin{figure}[t]
	\centering
\includegraphics[width=0.39\textwidth]{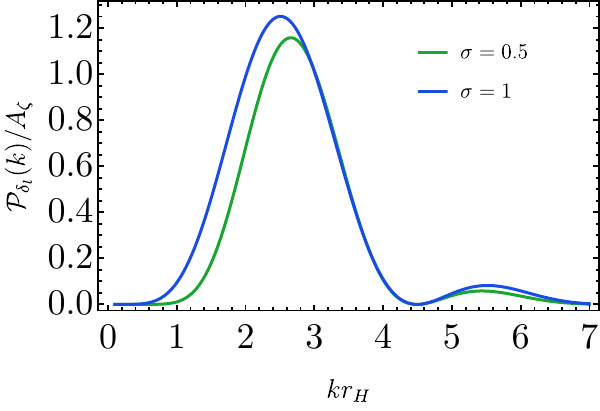}
	\caption{The integrand of the linear variance in Eq.~\eqref{sigma2} for a lognormal curvature perturbation power spectrum ${\cal P}_\zeta(k)= A_\zeta \, {\rm exp}\left[-\ln (2k/3k_\star)^2/2 \sigma^2 \right]$, assuming $k_\star r_m = 2.1 \, (1.75)$ for different widths $\sigma = 0.5 \, (1)$.
	}
	\label{linearvariance}
\end{figure}

\vskip 0.5cm
\noindent
\noindent{{\bf{\it A critical look at the  standard PBH abundance calculation.}}}
At this stage one can outline a few inconsistencies in the standard calculation of the PBH abundance:  

{\it i)} The  initial density perturbation is defined on superhorizon scales, and then evolved forward through horizon re-entry to check  if  a PBH will form. 
The starting point, Eq.~\eqref{deltaNL}, to calculate the threshold   is  rigorously valid only on superhorizon scales, much before the
large overdensities re-enter the Hubble radius.  The threshold is computed by using the time-independent part of the profile at times such that $r_m\gg r_H$, and then extrapolating its value at horizon crossing, for which $\epsilon(t_H)=1$, with  the growth in time $\sim(a H)^{-2}\sim a^2$. This procedure based on the gradient expansion breaks down close to  horizon crossing and neglects the effect of the radiation pressure, which would be increasingly more important as we approach horizon crossing. This is clear  from the expression (\ref{sh}) where there is no radiation transfer function.

The point  is that the variance $\sigma_l^2$, which contains the information of the radiation pressure through the linear radiation transfer function, 
 is obtained by integrating over all momenta and the integrand in its  definition, Eq.~\eqref{sigma2}, is peaked at scales beneath the horizon. Fig.~\ref{linearvariance} shows this point for some representative examples of comoving curvature perturbation power spectra. 

This means that the variance gets contributions from scales which at  $\epsilon(t_H)=1$ are already well inside the horizon. 
The action of the radiation pressure  becomes  so important that the density contrast stops growing, thus  changing completely the time dependence with respect to the linearly extrapolated case with no radiation pressure. This is because, for the relevant scales, the time dependence in the square root of the variance is of the form (up to oscillations) $\sim(a H)^{-2}T(k/aH)\sim$ constant.

The comparison between the  extrapolated threshold with no transfer function and $\sigma_l$, which determines the PBH abundance, is therefore not much consistent. Even at the linear level the time dependence between the threshold, which is 
computed considering only cosmic expansion and without pressure effects at horizon crossing,
and the square root of the variance does not  cancel in the standard procedure. 

One could of course compute both the density contrast and the square root of the variance on superhorizon scales to cancel the time dependence.
This would require  computing the  variance without including the
radiation transfer function~\cite{Young:2019osy,Yoo:2018kvb,Yoo:2020dkz}.  However, the  integral over the momenta in the variance would be sensitive to the ultraviolet cut-off due to the $\sim k^4$ term and therefore again sensitive to the details around horizon crossing.

 Forgetting the issue introduced by the radiation pressure and accounted for by the radiation transfer function does not come without a price already at the linear level. Indeed, calling $k_\star$ the typical momentum at which PBHs form, one has typically at the linear level $k_\star r_m \simeq k_\star/aH\simeq {\cal O}(2\divisionsymbol 3)$~\cite{Germani:2018jgr,Musco:2018rwt}\footnote{For a monochromatic curvature spectrum peaked at a given momentum, $k_\star$ coincides with such a momentum and $k_\star r_m\simeq 2.7$, while for a broad spectrum it coincides with the maximum momentum scale, as the PBH mass function peaks at that scale~\cite{DeLuca:2020ioi}, and $k_\star r_m\simeq 3.5$.}, and $T^2(k_\star r_m)\simeq (0.76\divisionsymbol 0.53)$. The effect of the transfer function is not negligible because the length scale $\sim k_\star^{-1}$ is already well inside the horizon when $r_m\simeq r_H$. The radiation pressure therefore tends to suppress the variance and makes it more difficult to build up perturbations able to overcome a given barrier. As we shall see, this remains true at second-order.

{\it ii)}  The quantity $\delta_m$  is intrinsically of the second-order. Evaluating the effects of the radiation pressure   will therefore require at least a   second-order transfer function  to be consistent both for the calculation of the threshold and the variance. This is a major source of uncertainty in the calculation of the PBH abundance. 
In the following we will adopt a perturbative approach to discuss the role of non-linearities in the evolution of the density contrast by expanding each quantity as $\mathcal{O} (t,{\bf x}) = \mathcal{O}_1 (t,{\bf x}) + \mathcal{O}_2 (t,{\bf x})/2 + \cdots$ (see the Appendix for the perturbative expansion of the cosmological perturbations).\footnote{This perturbative scheme will apply as well to the treatment of $r_m e^{\zeta(r_m)}$ in the definition of $\epsilon(t_H)$, such that at the linear level $r_m \simeq r_H$ at the time of horizon crossing.}

Consider  the density contrast  in the comoving gauge of Eq.~\eqref{deltaNL} close to the Hubble crossing time $t_H$ and imagine an effective expansion  at second-order of the form
\begin{align}
\delta^\com(t_H,{\bf x})&=f_\zeta(t_H)\nabla^2\zeta_1({\bf x})\nonumber\\
&+g_\zeta(t_H)\zeta_1({\bf x})\nabla^2\zeta_1({\bf x})\nonumber\\
&+h_\zeta(t_H)\partial_i\zeta_1({\bf x})\partial^i \zeta_1({\bf x})+\cdots,
\end{align}
in terms of some time-dependent functions $f_\zeta(t_H)$, $g_\zeta(t_H)$, $h_\zeta(t_H)$, $\cdots$.
The variance of such a quantity will depend only  (assuming $\zeta_1({\bf x})$ is a Gaussian field) upon   the square root of the combination 
\be
f_\zeta^2(t_H)\langle (\nabla^2\zeta_1)^2\rangle+g^2_\zeta(t_H)\langle(\zeta_1\nabla^2\zeta_1)^2\rangle+\cdots,
\ee
as $\zeta_1({\bf x})$ is treated as a stochastic quantity.  The threshold, on the other hand, is determined  from the profile in real space, and will depend on the linear combination of the functions $f_\zeta(t_H)$, $g_\zeta(t_H)$, $h_\zeta(t_H)$, $\cdots$. Since close to the horizon the functions are not expected to have the same time dependence, the cancellation operating at the linear level will be  in general no longer operative;

{\it iii)}  The threshold $\delta_{c}$ (or $\delta_{l,c}$) is routinely calculated extrapolating its value at horizon crossing in an unperturbed universe. However, it has been already noted that the non-linear effects arising at the true horizon crossing in an inhomogeneous universe, where the gradient expansion fails, increase the value of the threshold by about a factor ${\cal O}(1.7\divisionsymbol 2)$ with respect to the one computed on superhorizon scales and linearly extrapolated at $\epsilon(t_H)=1$~\cite{Musco:2020jjb} (see their Fig.~8). 
Part of this extra growth is due to the longer time
necessary to reach the non-linear horizon crossing,   part is
due to higher orders in the gradient expansion 
 when $\epsilon(t_H)\sim 1$ and finally part to non-linear effects due to the radiation pressure, which are accounted for numerically in Ref.~\cite{Musco:2020jjb}. In general, increasing the critical threshold by a factor ${\cal O}(1.7\divisionsymbol 2)$ might  drastically reduce the PBH abundance, unless the variance of the density contrast does increase correspondingly.

Overall, non-linear effects  become important close to horizon crossing and may give rise to large corrections to the probability of collapse estimated in the literature.  

The rest of the paper is dedicated to inspect more closely such sources of uncertainties due to   non-linear
effects. We start by commenting about the last  point we mentioned.

\vskip 0.5cm
\noindent
\noindent{{\bf{\it Crossing the real horizon later.}}}
There are various  effects to consider. First, the physical horizon in a perturbed universe expands more slowly than in an unperturbed universe, making the density profile grow more. Secondly,   non-linear effects and the effect of the radiation pressure close to the horizon crossing change the density contrast profile; both lead to an increase of the critical threshold, as discussed  numerically in Section 5 of Ref.~\cite{Musco:2020jjb}.

We discuss here the first effect, which  is already present at horizon crossing at the linear level, and we will come back to the second effect in the next section.

Consider the equation describing the expansion rate 
$H(t,{\bf x})$ as measured by a free-falling observer which is instantaneously at rest with respect to the radiation fluid, i.e. on comoving slicings \cite{Liddle:2000cg}
\be
\label{H}
H^2(t,{\bf x})=H^2(t)\Big(1+\delta_1^\com(t,{\bf x})\Big)+\frac{2}{3}
\frac{\nabla^2}{a^2}\zeta_1(t,{\bf x}).
\ee
Since for large amplitudes the peaks of the density contrast are identifiable with the peaks of the comoving curvature perturbation \cite{DeLuca:2019qsy} for which $\nabla^2\zeta_1<0$, we automatically deduce that the physical Hubble rate is smaller than the background value provided that the last  term in Eq.~\eqref{H} is larger than the term $H^2(t)\delta_1^\com$. One can  easily convince oneself that this is indeed the case.

At the linear level and for any scale (not only superhorizon), one can relate the linear density contrast in the comoving gauge to the linear gauge-invariant Bardeen's potential $\Phi_1(t,{\bf x})$ through the following equation \cite{Malik:2008im} (see also the Appendix):
\be
\delta_1^\com(t,{\bf x})=\frac{2}{3}\frac{\nabla^2\Phi_1(t,{\bf x})}{a^2H^2}.
\ee
Approximating at horizon crossing $\Phi_1(t_H,{\bf x})\simeq -2\zeta_1({\bf x})/3$, one obtains
\be
H^2(t_H,{\bf x})\simeq H^2(t_H)\left(1-\frac{1}{2}\delta_1^\com(t_H,{\bf x})\right).
\ee
One can estimate the effect on the threshold by averaging over a sphere of radius $r_m$ and assuming the critical profile for $\delta_1^\com(t_H,{\bf x})$, obtaining
\be
\label{l}
\frac{3}{r_m^3} \int_0^{r_m} \d r \, r^2 \lp \frac{H^2(t_H,{\bf x})}{H^2(t_H)} -1 \rp   \simeq -\frac{1}{2}\delta_c,
\ee
from which one can see already the effect of the curvature term in slowing down the Hubble radius expansion. In the last equation, to be consistent with the linear perturbation theory adopted so far, one needs to take the value of $\delta_c$ obtained from the  linear Gaussian threshold~\cite{Musco:2020jjb} for the volume average of the first-order density contrast $\delta_1^\com(t_H,{\bf x})$. 

One can do a better estimate  including the effect of higher-order gradients, still  remaining at the linear level.
At horizon crossing, the comoving curvature term starts growing due to the scalar shear 
and the effect of higher-order gradients starts becoming manifest \cite{Malik:2008im}
\be
\frac{\dot\zeta_1}{H}\simeq \frac{1}{4}\delta^\com_1
-\frac{3}{4}\frac{\nabla^2 \delta_1^\com}{a^2H^2},
\ee
where the dot denotes differentiation with respect to the cosmic time.
The rate  $\dot\zeta_1$ is positive since the second term  grows faster (as $a^2$) than the first one and around  the peak of the density contrast $\nabla^2\delta_1^\com<0$.
Integrating over time and inserting the result in Eq.~\eqref{H} gives
\begin{align}
 \lp \frac{H^2(t_H,{\bf x})}{H^2(t_H)} -1 \rp  &\simeq \delta_1^\text{\tiny com}+ \frac{1}{12} \frac{\nabla^2 \delta_1^\text{\tiny com}}{a^2 H^2} - \frac{1}{4} \frac{\nabla^4  \delta_1^\text{\tiny com}}{a^4 H^4}.
\end{align}
When the perturbation has just crossed the horizon, the last two terms dominate and  averaging again over a sphere of radius $r_m$ we obtain an extra shift in the Hubble rate
\begin{align}
& \frac{3}{r_m^3} \int_0^{r_m} \d r \, r^2 \lp \frac{H^2(t_H,{\bf x})}{H^2(t_H)} -1 \rp \nonumber \\
& \simeq \frac{7}{4}r_m {\delta_1'}^{\com}(r_m) - \frac{3}{2}r_m^2 {\delta_1''}^{\com}(r_m) - \frac{3}{4}r_m^3 {\delta_1'''}^{\com}(r_m) \nonumber\\
&\sim -
\frac{5}{2}\delta_1^{\com}(r_m)  
\sim  -\frac{5}{6}\delta_c,
\end{align}
where we have approximated $r_m{\delta_1'}^{\com}(r_m)\sim -r_m^2{\delta_1''}^{\com} (r_m) \sim r_m^3{\delta_1'''}^{\com} (r_m) \sim -{\delta_1^\com} (r_m)$ for a peaked perturbation,
and  used the relation $\delta_c = 3 \delta_1^\text{\tiny com} (t_H,r_m)$ for the critical profile~\cite{Musco:2018rwt}, to get an order of magnitude estimate.
Taking $\delta_c\simeq 0.51$ for the linear Gaussian threshold~\cite{Musco:2020jjb}, we obtain a change in the horizon-crossing time $\epsilon (t_H) \propto 1/H$ of
\be
\epsilon(t_H)
\simeq \frac{1}{\left(1-5\delta_c/6\right)^{1/2}}\simeq 1.3,
\ee
which is  already in good agreement with what was numerically found in Ref.~\cite{Musco:2020jjb} (see their Fig.~5, where a monochromatic spectrum corresponds formally to the case $\alpha=6.33$).

Being the  variance dominated by the momentum modes well inside the horizon at $r_m=r_H$, when the linear density contrast has already stopped growing, increasing $\epsilon(t_H)$ will not necessarily change the linear variance by the same factor. For instance, for a broad spectrum of the curvature perturbation, the linear variance does not change shifting $r_H$ as the integral is over the momenta $k r_H$ once $r_m$ is set to be equal to $r_H$. For a monochromatic power spectrum ${\cal P}_\zeta(k)=A_\zeta k_\star\delta(k-k_\star)$, the change of the variance is set by the square of the radiation transfer function and it does not automatically cancel the   change of $\epsilon^2(t_H)$ in the critical threshold.

We do not attempt here to calculate the  effect of non-linearities on the slowing down of horizon crossing, which have been studied numerically  in Ref.~\cite{Musco:2020jjb}. We limit ourselves to the observation  that one expects corrections of the order of ${\cal O}(\delta_c^2)$, see for instance Eq.~\eqref{eq:zeta_so_def} of the Appendix. For instance,   the leading (and growing in time)  second-order correction to the comoving curvature perturbation at horizon crossing  has the form~\cite{Bartolo:2003bz,Malik:2008im,Inomata} 
\be
\label{nonspher}
\frac{2}{3}\nabla^2\zeta_2(t,{\bf x})\simeq -\frac{1}{48a^2H^2}\sum_{i\neq j}\partial_i\partial_j\delta_1^\com(t,{\bf x})\partial^i\partial^j\delta_1^\com(t,{\bf x}),
\ee
which is again negative and increasing like $a^2$.   Its  effect is to  further slow down the expansion rate and increase the PBH threshold,  but  only for  non-spherical peaks.

\vskip 0.5cm
\noindent
\noindent{{\bf{\it The non-linear  transfer function: the second-order case.}}}
As we have stressed in the introduction, one of the main uncertainties in the calculation of the PBH abundance is that the effect of the radiation pressure is standardly either not accounted  for (in the threshold, with the exception of  Ref.~\cite{Musco:2020jjb})  or it is, but only at the linear level (in the variance).
We would like to offer some considerations about the non-linear effects, limiting ourselves to the second-order. 

Our starting point is the second-order equation for the density contrast in the comoving orthogonal gauge expressed in terms of the  Bardeen potentials constructed from the longitudinal gauge (see the Appendix)
\begin{align}
\label{fund}
 \frac{3}{2}a^2H^2\delta_2^\com &=  \nabla^2 \Psi_2
 + 3\partial_i \Phi_1\,\partial^i \Phi_1
+ 8\Phi_1 \nabla^2 \Phi_1\nonumber\\
 &+3\dot \Phi_1^2 
 +\frac{6H}{\nabla^2}\left(\partial^i\dot\Phi_1\partial_i{\Phi_1}+\dot\Phi_1\nabla^2{\Phi_1}\right).
\end{align}
This expression is valid at all scales, not only on superhorizon. It shows a fundamental point, that close to horizon crossing the overdensity on  comoving orthogonal slices contains many   non-linear terms, even non-local, and therefore it is unlikely that
the non-linear smoothed density contrast will have a quadratic relation of the local type with the linear one, as in Eq.~\eqref{deltam}. In other words, Eq.~\eqref{deltam}, obtained on superhorizon scales,  fails to capture the non-linearities
in the critical threshold due to the radiation pressure. This  implies that  neither $\delta_c$ nor $\delta_{l,c}$ may be used and that a priori there is no simple relation of the local type between the linear density contrast and the  non-linear  one.

Eq.~\eqref{fund} needs to be supplemented with the equation for the evolution of the second-order gravitational potential $\Psi_2$~\cite{Bartolo:2006fj,Inomata}
\begin{align}
\label{psi2text}
{\ddot \Psi_2}+5 H{\dot \Psi_2}-\frac{1}{3}\frac{\nabla^2\Psi_2}{a^2}&=\frac{1}{a^2}\left[\frac{2}{3} \partial^i \Phi_1 \partial_i \Phi_1  + \frac{8}{3}\Phi_1\nabla^2\Phi_1 \right.\nonumber\\
&
+2 {\dot \Phi_1}^2 +H N^j_{\,\,\, i}\left(A^{(2)i}_{\,\,j}\right)^\cdot\nonumber\\
&+\left.\frac{1}{3}N^j_{\,\,\, i}\left(A^{(2)i}_{\,\,j}\right)^{,k}_{,k}\right]\\
A^{(2)i}_{\,\,j}&=6\,\partial^i\Phi_1\partial_j\Phi_1+\frac{2}{H}\left(\partial^i\Phi_1\partial_j\Phi_1\right)^\cdot\nonumber\\
&+\frac{2}{H^2}\partial^i\dot \Phi_1\partial_j\dot\Phi_1,\nonumber
\end{align}
where
\be
N^j_{\,\,\, i}=\frac{3}{2}\frac{1}{\nabla^2}
\left(\frac{\partial^j\partial_i}{\nabla^2}
-\frac{1}{3}\delta^j_{\,\,\, i}\right).
\ee
Eq.~\eqref{psi2text} has to be solved with the superhorizon initial condition~\cite{Bartolo:2006cu,Bartolo:2006fj}
\be
\label{initial}
{\Psi_2}({\bf x})=
-2\Phi^2_1({\bf x})+2N^j_{\,\,\,i}\left(\partial^i\Phi_1({\bf x})\partial_j\Phi_1({\bf x})\right).
\ee
In momentum space the solution is the sum of two pieces
\begin{align}
\label{psi2Ipsi}
&{\Psi_2}(t,{\bf k})={\Psi_2}({\bf k})T(k/aH) \nonumber \\ 
& +\int\frac{{\rm d}^3p}{(2\pi)^3}u v \left(\frac{2}{3}\right)^2  I_\Psi \lp u,v,\frac{k}{a H}\rp \zeta_1(\bf p)\zeta_1({\bf k}-{\bf p}),
\end{align}
where $u=|{\bf k}-{\bf p}|/k$, $v=p/k$. This equation shows the dependence on the linear radiation transfer function in the first term and of the 
second-order radiation transfer function through
the function $I_\Psi(u,v,x)$~\cite{Bartolo:2006cu,Bartolo:2006fj,Inomata}.

To account for the effect of the radiation pressure in the PBH threshold  at second-order in perturbation theory we proceed as follows. At first-order the equation of motion of the gravitational potential is 
\be
{\ddot \Phi_1}+5 H{\dot \Phi_1}-\frac{1}{3}\frac{\nabla^2\Phi_1}{a^2}=0.
\ee
We can treat the last term as a perturbation, which gives close to horizon crossing
\begin{align}
\Phi_1(t,{\bf x})&=
\int^t\frac{{\rm d} t'}{a^5(t')}\int^{t'}{\rm d} t''a^5(t'') \frac{1}{3}\frac{\nabla^2\Phi_1({\bf x})}{a^2(t'')}\nonumber\\
&=
\left(1+\frac{1}{30}\frac{\nabla^2}{a^2H^2}\right)\Phi_1({\bf x}).
\end{align}
This is quite a good approximation:  going to momentum space and evaluating at $k_\star r_m \simeq k_\star r_H=2.74$ for a monochromatic power spectrum, we find $\Phi_1(t,{\bf x})\simeq 0.75 \Phi_1({\bf x})$,  while the linear radiation transfer function gives
$T(2.74) \Phi_1({\bf x}) =0.77 \Phi_1({\bf x})$.

At second-order, proceeding in a similar way to obtain $\Psi_2 (t,{\bf x})$ by using Eq.~\eqref{initial}, then shown in Eq.~\eqref{solPsi2threshold}, we obtain (see the Appendix for details)
\begin{align}
\label{delta2com}
 \delta_2^\com  &\simeq  \frac{1}{a^2H^2} \lp \frac{2}{3} \nabla^2 \Psi_2
 + 2\partial_i \Phi_1\,\partial^i \Phi_1
+ \frac{16}{3}\Phi_1 \nabla^2 \Phi_1 \rp \nonumber\\
& = \frac{1}{a^2H^2} \lp - \frac{4}{3} \nabla^2 \Phi^2_1({\bf x}) + 2 \frac{\partial^j\partial_i}{\nabla^2} (\partial^i\Phi_1({\bf x})\partial_j\Phi_1({\bf x})) \right. \nonumber \\
& \left. + \frac{4}{3} \partial^i \Phi_1({\bf x}) \partial_i \Phi_1({\bf x}) +  \frac{16}{3}\Phi_1 \nabla^2 \Phi_1 \rp \nonumber \\ 
& + \frac{1}{a^4 H^4} \lp 
\frac{2}{3} \partial^i \nabla^2 \Phi_1({\bf x}) \partial_i \Phi_1({\bf x}) \right. \nonumber \\
& \left.
+ \frac{8}{15} \nabla^2 \Phi_1({\bf x}) \nabla^2 \Phi_1({\bf x})   + \frac{2}{15} \Phi_1({\bf x}) \nabla^4  \Phi_1({\bf x})
\rp.
\end{align}
When evaluated at the time of horizon crossing $a H = 1/r_H \simeq 1/r_m$, the first-order curvature profile $\Phi_1(t_H,{\bf x})\simeq -2\zeta_1({\bf x})/3$ in Fourier space is dominated by large momenta, see for instance Fig.~\ref{linearvariance}  for different shapes of a lognormal power spectrum.  This allows us to focus only on the last three lines in the previous expression, which are expected to dominate for large momenta. 

\begin{figure}[t]
	\centering
\includegraphics[width=0.39\textwidth]{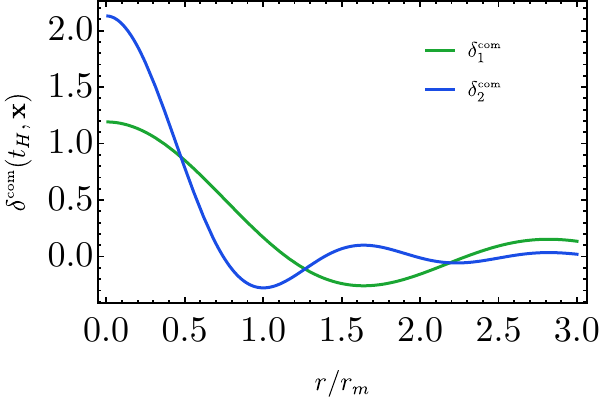}
	\caption{The density profile corresponding to the critical threshold at first- and second-order in perturbation theory, for a monochromatic curvature spectrum. 
	}
	\label{densities}
\end{figure}

At horizon crossing the density contrast becomes
\begin{align}
\label{delta2threshold}
 \delta_2^\com (t_H,{\bf x})  &\simeq  \frac{4r_m^4}{9} \lp
\frac{2}{3} \partial^i \nabla^2  \zeta_1({\bf x}) \partial_i \zeta_1({\bf x}) 
\right. \nonumber \\
& \left.
+ \frac{8}{15} \nabla^2 \zeta_1({\bf x}) \nabla^2 \zeta_1({\bf x})
 +  \frac{2}{15} \zeta_1({\bf x}) \nabla^4  \zeta_1({\bf x})
\rp.
\end{align}
Adopting Gaussian peak theory to a monochromatic curvature perturbation power spectrum, for which 
 $k_\star r_m \simeq 2.74$, 
one can derive the shape for the linear curvature profile~\cite{Musco:2020jjb} 
\be
\zeta_1({\bf x}) = \zeta_0 \frac{\sin (k_\star r)}{k_\star r},
\ee
from which we obtain the critical density profile at second-order shown in Fig.~\ref{densities}, where the overall amplitude $\zeta_0$ has been fixed at the linear level by requiring that $\delta_c = 0.51 = 3 \delta_1^\text{\tiny com} (t_H,r_m)$~\cite{Musco:2018rwt}.
The impact of the second-order radiation transfer function  on the PBH threshold is estimated to be
\be
 \frac{3}{r_m^3} \int_0^{r_m} \d r r^2 \delta_2^\com (t_H, r)\simeq 0.13.
\ee
This result shows that the second-order radiation transfer function  increases further the PBH threshold for a monochromatic spectrum, due to the radiation pressure which makes the collapse of the overdensities into a PBH  more difficult.  
To this effect one should further add the one described in the previous section related to the slowing down of the horizon. 
The corresponding critical threshold will be given roughly by (the factor $1/2$ comes from the normalization of the second-order variables)
\be
\epsilon^2(t_H)\delta_c + \frac{1}{2}\epsilon^4(t_H) \frac{3}{r_m^3} \int_0^{r_m} \d r r^2 \delta_2^\com \simeq 1.08,
\ee
where the corrective factor $\epsilon^4(t_H)$ has been included for the slowing down of the horizon based on the dependence $1/(a H)^4$ in Eq.~\eqref{delta2threshold}.
This result is in good agreement with the numerical fit for the PBH threshold, $\alpha^{0.06} + 0.025 \simeq 1.14$, for a monochromatic curvature power spectrum with $\alpha = 6.33$, provided in  Ref.~\cite{Musco:2020jjb} (see also the left panel of their Fig. 8).

\begin{figure}[t]
	\centering
\includegraphics[width=0.39\textwidth]{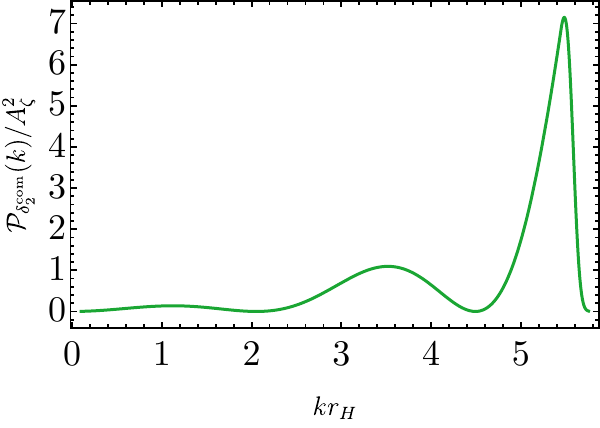}
	\caption{The integrand of the second-order variance in Eq.~\eqref{integrand2ndmono} for a monochromatic curvature perturbation power spectrum evaluated at the time of horizon crossing. 
	}
	\label{2ndvariance}
\end{figure}

As for the variance at second-order, we offer the following considerations. When a given scale re-enters the horizon, the linear gravitational potential $\Phi_1(t,{\bf x})$ decays like $a^{-2}$ (up to oscillations). By inspecting the time dependence of Eq.~\eqref{psi2Ipsi} it is easy to convince ourselves that the second-order gravitational potential
$\Psi_2(t,{\bf x})$ decays like $a^{-2}$ as well~\cite{Inomata}. As a consequence, one can keep the leading time dependent term in Eq.~\eqref{delta2com} and approximate
\be
 \frac{3}{2}a^2H^2\delta_2^\com \simeq  \nabla^2 \Psi_2,
\ee
because all the other terms decay faster at horizon crossing. From the solution of the second-order scalar perturbation in Eq.~\eqref{psi2Ipsi}, one can estimate the variance of the second-order smoothed  density contrast in the comoving gauge at horizon crossing by computing its two-point connected correlation function, which gives
\begin{align}
\sigma_2^2 &= \int\frac{{\rm d}k}{k} {\cal P}_{\delta_2^\com}(k, r_H),\nonumber\\
{\cal P}_{\delta_2^\com}(k,r_H)&=\int_0^\infty {\rm d}v\int_{|v-1|}^{v+1}{\rm d}u \left(\frac{2}{3}\right)^6
(k r_H)^4  I^2_\Psi(u,v,k r_H) \nonumber \\
& \cdot W^2(k,r_m) {\mathcal P}_{\zeta_1}(kv) {\mathcal P}_{\zeta_1}(ku).
\end{align}
In Fig.~\ref{2ndvariance} we plot the power spectrum of the second-order
smoothed density contrast in the comoving gauge for a monochromatic curvature perturbation power spectrum,  
for which we get 
\begin{align}
\label{integrand2ndmono}
{\cal P}_{\delta_2^\com}(k,r_H)&=\left(\frac{2}{3}\right)^6(k r_H)^4 \left(\frac{k_\star}{k}\right)^2 W^2(k_\star,r_m) \nonumber\\
&\cdot I^2_\Psi\left(\frac{k_\star}{k},\frac{k_\star}{k},k_\star r_H\right) \theta \lp 2-\frac{k}{k_\star}\rp,
\end{align}
evaluated for $k_\star r_m \simeq k_\star r_H=2.74$, where $\theta$ indicates the Heaviside step function. The figure confirms that, as in the linear case, the variance is dominated by momenta larger than the Hubble rate where the second-order radiation transfer function  is non-negligible. At this stage the time dependence $\sim a^{-2}$ of the gravitational  potential $\Psi_2$ is the same as at the linear level and the second-order variance  does not change as well with time (up to oscillations). The values of the linear and second-order variances
 for $k_\star r_m = 2.74$ are
 (the factor $1/4$ in the second-order term accounts for the normalization of the second-order density contrast)
\be
\sigma_l^2\simeq 1.19\, (0.41)\, A_\zeta\,\,\,{\rm and}{\,\,\,} \frac{1}{4}\sigma_2^2\simeq 0.25\, (1.02)\, A_\zeta^2,
\ee
where in the parenthesis we have indicated the values when the Hubble rate is shifted by a factor $\epsilon(t_H) \simeq 1.3$ due to the crossing of the physical horizon, as discussed in the previous section.
We deduce that the second-order  contribution is always negligible compared to the linear one, even for the preferred value of $A_\zeta\simeq 6\cdot 10^{-2}$ to explain the recent PTA data~\cite{Franciolini:2023pbf}. We think this  conclusion is quite robust and does not depend on the details of the time of horizon crossing and the shape  of the comoving curvature power spectrum, as long as $k_\star r_m\gsim 1$. For instance, for a broad spectrum, even at second-order the variance is insensitive to changes in the physical horizon as it is an integral over all the momenta $k r_H$ once one sets  $r_m=r_H$ at the corresponding order in perturbation theory.

As a final remark we notice that, because of the non-Gaussian nature of the second-order density contrast $\delta_2^\com$, the expression for the PBH probability will depend also on higher-order cumulants of the density field beyond the two-point function.
We leave the investigation of the form of such a probability to future work. 


\vskip 0.5cm
\noindent
\noindent{{\bf{\it Conclusions.}}}
In this article we have addressed various issues related to the calculation of the abundance of PBHs and point out the uncertainties the standard procedure suffers of. They come above all from the incomplete treatment of the radiation pressure  and of the non-linear effects close to horizon crossing. 

Even if our calculations are certainly incomplete, we can offer the following hints: {\it i)} the critical threshold is larger than what routinely assumed. We have provided some physical explanations for such an increase; {\it ii)} The variance of the non-linear density contrast is unlikely to be changed with respect to the linear one. This is because it is suppressed by an extra power of the curvature perturbation power spectrum; {\it iii)} The time dependence between the critical threshold   and the square root of the variance does not cancel at horizon crossing as the latter is dominated by momenta larger than the horizon, when the density contrast has already frozen  in (up to oscillations). This is because the characteristic length scale is such that  $k_\star^{-1}\gsim r_m\simeq r_H$. The reason why the linear and non-linear variances are suppressed is because of the action of the radiation pressure at the scales close to $k_\star$. Therefore, it is difficult to build up fluctuations which overcome a given barrier; {\it iv)} The relation between the non-linear and linear threshold is not of the local type, therefore making the calculation of the formation probability not straightforward; {\it v)} Given the fact that the non-linear effects increase the critical threshold, but not the variance, it is likely that the PBH abundance is much smaller than what considered in the literature so far. 

If confirmed by a more complete  all-order (most likely numerical) calculation,  such conclusions would  make the PTA data compatible with the PBH scenario, since the same amplitude of the stochastic GW background would correspond to much smaller PBH abundances.

\vskip 0.5cm
\noindent
\noindent{{\bf{\it Acknowledgments.}}}
We thank  I. Musco for several discussions about the procedure to calculate the critical threshold and I. Musco and G. Franciolini for  comments on the draft. V.DL. is supported by funds provided by the Center for Particle Cosmology at the University of Pennsylvania. 
A.R. is supported by the Boninchi Foundation for the project ``PBHs in the Era of GW Astronomy''.

\bibliography{draft}
\newpage

\onecolumngrid
\begin{center}
{\bf APPENDIX}
\end{center}
\renewcommand{\theequation}{A.\arabic{equation}}
\setcounter{equation}{0}
\label{intro}
\noindent
The components of a perturbed spatially flat Robertson-Walker 
metric can be written, up to second-order and using conformal time, as (primes indicate here differentiation with respect to the conformal time $\tau$)~\cite{Bartolo:2006fj}
\begin{align} \label{metric1}
g_{00}&=-a^2(\tau)\left( 1+2 \phi_1+\phi_2 \right),\nonumber\\
g_{0i}&=a^2(\tau)\left( \hat{\omega}_{1i}+\frac{1}{2} 
\hat{\omega}_{2i} \right)
,  \nonumber\\g_{ij}&=a^2(\tau)\left[
(1 -2 \psi_1 - \psi_2)\delta_{ij}+
\left( \hat{\chi}_{1ij}+\frac{1}{2}\hat{\chi}_{2ij} \right)\right].
\end{align}
The standard splitting of the perturbations into scalar, transverse 
({\it i.e} divergence-free) vector, and transverse trace-free tensor 
parts, with respect to the three-dimensional space with metric $\delta_{ij}$, 
can be performed in the following way:
\begin{align}
\hat{\omega}_i &=\partial_i\omega+\omega_i, \nonumber \\
\hat{\chi}_{ij}&=D_{ij}\chi+\partial_i\chi_j+\partial_j\chi_i
+\chi_{ij},
\end{align}
where  $\omega_i$
and $\chi_i$ are transverse vectors ($\partial^i\omega_i=
\partial^i\chi_i=0$), $\chi_{ij}$ is a symmetric transverse and 
trace-free tensor ($\partial^i\chi_{ij}=0$, $\chi_i^{~i}
=0$) and $D_{ij}=\partial_i \partial_j - (1/3) \, \, 
\delta_{ij}\, \partial^k\partial_k$ is a trace-free operator. Here and in the following latin indices 
are raised and lowered using $\delta^{ij}$ and $\delta_{ij}$, respectively.
For our purposes the metric in Eq.~\eqref{metric1} can be simplified. In fact, 
first-order vector perturbations are not generated during inflation and tensor modes are 
 negligible.
Thus, in the following we can neglect 
$\omega_{1i}$, $\chi_{1i}$ and $\chi_{1ij}$.
However the same is not true for the second-order perturbations. 
In the second-order theory the second-order vector and tensor 
contributions can be generated by the first-order scalar perturbations 
even if they are initially zero. Thus we have to take them into account and we shall use the metric
\begin{align} \label{metric2}
g_{00}&=-a^2(\tau)\left( 1+2 \phi_1+\phi_2 \right),\nonumber\\
g_{0i}&=a^2(\tau)\left( \partial_i\omega_1+\frac{1}{2}\, 
\partial_i\omega_2+\frac{1}{2}\, \omega_{2i} \right)
,  \nonumber\\g_{ij}&=a^2(\tau)\left[
\left( 1 -2 \psi_1 - \psi_2 \right)\delta_{ij}+
D_{ij}\left( \chi_1 +\frac{1}{2} \chi_2 \right)+\frac{1}{2}\left( \partial_i\chi_{2j}
+\partial_j\chi_{2i}
+\chi_{2ij}\right)\right].
\end{align}
The first-order perturbations of the Einstein tensor components which we  need are
\begin{align}
\label{100}
\deu {G_{~0}^{0}}&= \frac{1}{a^2}\Bigg[ 6\,\Ab \phi_1 \,+\,6\,\Aa{\psi'_1}\,+\, 2\,\Aa \La
 \omega_1 - 2\,\La \psi_1 \,-\, \frac{1}{2}\,\partial_k \partial^i
 \,D^k_i \chi_1 \Bigg] , \\
\label{10i}\deu {G^0_{~i}}&= \frac{1}{a^2}\left( -\,2\,\Aa \partial_i \phi_1 \,-\, 2\,\partial_i
{\psi'_1} \,-\, \frac{1}{2}\,\partial_k D^k_{~i}{\chi'_1}\right).
\end{align}  
At second-order they are 
\begin{align}
\label{200}
\ded {G_{~0}^0} &= \frac{1}{a^2} \Big[ 3\Ab \phi_2\,+\,3\,\Aa {\psi'_2} \,
-\, \La \psi_2 +\Aa \La \omega_2 \,-\,\frac{1}{4} \partial_k \partial_i\,D^{ki} \chi_2
-\, 12 \left( \Aa \right)^2 \left( \phi_1 \right)^2 \nonumber\\
&- 12\,\Aa\,\phi_1\,{\psi'_1} 
-\,3\,\partial_i \psi_1\,\partial^i \psi_1 \,-\, 8\,\psi_1\ \La \psi_1- 3 \left( {\psi'_1}\right)^2 \,+\, 4\,\Aa\,\phi_1\,\La \omega_1\,-\, 2\,\Aa
\partial_k \omega_1\,\partial^k \phi_1 \nonumber\\
&-
\frac{1}{2} \Ac\,\partial_k \omega_1\,\partial^k \omega_1+\, 12\,\Aa \psi_1\,{\psi'_1} + \frac{1}{2}\,\partial_i\partial_k \omega_1\,\partial^i\partial^k \omega_1
-\,2\,\Aa \partial_k \psi_1\,\partial^k \omega_1 \nonumber\\
&+\, 4\,\Aa \psi_1\,\La
\omega_1\,-\, \frac{1}{2}\,\partial_k\partial^k \omega_1\,\partial_k\partial^k \omega_1 - 2\, \partial_k \omega_1 \,\partial^k {\psi'_1} \,-\,
2{\psi'_1}\La \omega_1 \,-\, \phi_1\,\partial_i\partial^k \,D^i_{~k} \chi_1\nonumber\\
&-\,
2\,\psi_1
\partial_k\partial^i \,D^k_{~i} \chi_1 
+ \partial_k \partial_i \psi_1\,D^{ki} \chi_1 - 2 \Aa
\partial_i\partial_k \omega_1\,D^{ik} \chi_1 \,-\,2\,\Aa \partial_k \omega_1 \,\partial_i\,
D^{ik} \chi_1\nonumber\\
&-\,\partial_k \omega_1\,\partial^i D^k_{~i} {\chi'_1}
- \frac{1}{2}\,\La \,D_{mk} \chi_1 \, D^{km}\chi_1 +
\partial_m\partial^k \,D_{ik} \chi_1\,D^{im} \chi_1 \,+\,
\frac{1}{2} \,\partial_k D^{km} \chi_1\, \partial^i D_{mi} \chi_1 \nonumber \\
&- \frac{1}{8}\,\partial^i D^{km}\chi_1 \, \partial_i D_{km} \chi_1 \,+\,
\frac{1}{8}\,D^{ik} {\chi'_1} \,D_{ki} {\chi'_1} + \Aa
D^{ki} \chi_1 \, D_{ik} {\chi'_1}\Big], 
\end{align}
\begin{align}
\label{20i}
\ded {G^0_{~i}} &= \frac{1}{a^2} \Big( - \Aa \partial_i \phi_2 \,-\,\partial_i
{\psi'_2} \,-\, \frac{1}{4}
\partial_k\,D^k_{~i} {\chi'_2} + \frac{1}{4} \partial_k\partial^k  \omega_{2i} \,+8 \Aa \phi_1 \partial_i \phi_1 
+\,4\,\phi_1\,\partial_i {\psi'_1}
+ 2 \,{\psi'_1}\partial_i \phi_1\nonumber\\
&-\, 4\,{\psi'_1}\partial_i
\psi_1 \,-\, 4\,\psi_1\,\partial_i
 {\psi'_1} \,+\, \partial_i \phi_1\,\La \omega_1 - \partial_i\partial_k \omega_1 
\,\partial^k \phi_1 
+8 \Ac \phi_1\,\partial_i \omega_1 \,-\, 4 \Ab \phi_1\,\partial_i \omega_1\nonumber\\
 &-\, 2\,\Aa \,\partial^k \omega_1\,\partial_i\partial_k \omega_1
  \,+\,\La \psi_1\,\partial_i \omega_1 + \partial^k \omega_1\,\,\partial_i\partial_k \psi_1
 \,-\,\partial_k \phi_1\,D^k_{~i} {\chi'_1}\,+\,
 \frac{1}{2}\,\partial^k \phi_1 \,D_{ik} {\chi'_1} \nonumber\\
 &-
 \psi_1\,\partial_k D^k_{~i} {\chi'_1} 
+\,
 \frac{1}{2}\,\partial_k\psi_1 \,D^k_{~i} {\chi'_1}
 \,-\, {\psi'_1} \partial_k D^k_{~i} \chi_1 -\partial_k{\psi'_1} D^k_{~i} 
\chi_1
 \,+\,\partial_i \omega_1\, \partial_k
 \partial^m\,D^k_{~m} \chi_1  
\nonumber\\
 &+\,\Ab \partial^k \omega_1\,D_{ik} \chi_1  \,+\, \partial^k \omega_1\,\partial_m\partial_i 
\,D^m_{~k} \chi_1 - \frac{1}{2}\,\partial^m \omega_1 \,\partial_k \partial^k
 \,D_{im} \chi_1
  \,+\, \frac{1}{2}\,\partial_k D^{km}\chi_1 \,D_{im} {\chi'_1}\nonumber\\
 &+\,\frac{1}{2} \, \partial_k D_{im} {\chi'_1}\,D^{km} \chi_1 
-
 \frac{1}{4} \partial_i D_{mk} \chi_1 \,D^{km} {\chi'_1} - \frac{1}{2}\, \partial_i D_{mk} {\chi'_1}\,D^{km} \chi_1- 2 \,\Ac\, \partial^k \omega_1\,D_{ik} \chi_1 \Big).
 \end{align}
 Correspondingly, the first-order components
 of the energy-momentum tensor are
\begin{align}
\deu {T_{~0}^{0}}&=-\delta\rho_1\, \nonumber\\
\deu {T_{~i}^{0}}&=(\rho_0+P_0)(v_{1i}+\hat{\omega}_{1i}),
\end{align}
in terms of the background energy density $\rho_0$ and pressure $P_0$.
At second-order they read
\begin{align}
\ded{T_{~0}^{0}}&=-\frac{1}{2}\delta\rho_2-(\rho_0+ P_0)v_{1k}(v_1^{k}+\hat{\omega}_1^{k}),\\
\ded{T_{~i}^{0}}&=\frac{1}{2}(\rho_0+P_0)\left[(v_{2i}+\hat{\omega}_{2i}-2\phi_1(v_{1i}+2\hat{\omega}_{1i})+ 4\deu g_{ik}v_1^{k}\right]+\frac{1}{2}(\delta\rho_1+\delta P_1)(v_{1i}+\hat{\omega}_{1i}).
\end{align}
Einstein's equations, as written in a generic gauge, are automatically gauge invariant and one can check that they can be written in terms of gauge-invariant quantities. One is therefore free to evaluate them in the most convenient gauge on the left- and on the right-hand side.
For instance, we evaluate the first-order Einstein tensor in the generalized longitudinal gauge (also called Poisson gauge) for which $\hat{\omega}=\chi=0$, while we evaluate the energy-momentum tensor in the comoving orthogonal gauge for which $v_i=\hat{\omega}_i=0$.
In such a case, one easily finds
by subtracting  Eqs.~\eqref{100} and \eqref{10i}, 
\be
\nabla^2\Phi_1=\frac{3}{2}{\cal H}^2\delta_{1}^\com,
\ee
where $\Psi_1=\Phi_1$ are the Bardeen's gauge-invariant potentials~\cite{Bardeen:1980kt}, ${\cal H} = a^\prime/a = a H$, and $\delta_1^\text{\tiny com}$ is the gauge-invariant density contrast in the comoving orthogonal gauge. Using the fact that in momentum space
\be
\Phi_1(k,\tau)=-\frac{2}{3}T(k\tau)\zeta_1(k),
\ee
 we get
\be
\delta_1^\text{\tiny com}(k,\tau)=\frac{4}{9}\frac{k^2}{{\cal H}^2}T(k\tau)\zeta_1(k).
\ee
This equation is valid at all scales, not only on superhorizon scales.
Similarly, at second-order one obtains the equations
\begin{align}
  -\frac{3}{2}{\cal H}^2\delta_2^\text{\tiny com} &= 3{\cal H}^2 \Phi_2\,+\,3\,{\cal H} {\Psi'_2} \,
-\, \La \Psi_2- 12 {\cal H}^2 \left( \Phi_1 \right)^2 
-3\,\partial_i \Phi_1\,\partial^i \Phi_1
\,-\, 8\,\Phi_1\ \La \Phi_1 -
 3\,\left( {\Phi'_1} \right)^2 , \nonumber \\
0 &=  - \,{\cal H} \partial_i \Phi_2 \,-\,\partial_i
{\Psi'_2}  \,+8 {\cal H} \Phi_1 \partial_i \Phi_1 - 2 \,{\Phi'_1}\partial_i \Phi_1,
 \end{align}
which give
\begin{align}
 \frac{3}{2}{\cal H}^2\delta_2^\text{\tiny com} &=  \nabla^2 \Psi_2
 + 3\partial_i \Phi_1\,\partial^i \Phi_1
+ 8\Phi_1 \nabla^2 \Phi_1+
 3\left( {\Phi'_1} \right)^2 
 +\frac{6{\cal H}}{\nabla^2}\left(\partial^i{\Phi'_1}\partial_i{\Phi_1}+{\Phi_1}^\prime\nabla^2{\Phi_1}\right),
\end{align}
which reproduces Eq.~\eqref{fund} of the main text.
The equation of motion for the second-order gravitational potential can be found in Refs.~\cite{Bartolo:2006fj,Inomata}
and during the radiation phase it reads
\begin{align}
\label{psi2}
{\Psi''_2}+4{\cal H}{\Psi'_2}-\frac{1}{3}\nabla^2{\Psi_2}&=\frac{2}{3} \partial^i \Phi_1 \partial_i \Phi_1 + \frac{8}{3}\Phi_1\nabla^2\Phi_1
+2\left( {\Phi'_1} \right)^2 +{\cal H}N^j_{\,\,\, i}\left(A^{i}_{2\,\,j}\right)^\prime+\frac{1}{3}N^j_{\,\,\, i}\left(A^{i}_{2\,\,j}\right)^{,k}_{,k}, \\
A^{i}_{2\,\,j}&=6\,\partial^i\Phi_1\partial_j\Phi_1+\frac{2}{{\cal H}}\left(\partial^i\Phi_1\partial_j\Phi_1\right)^\prime
+\frac{2}{{\cal H}^2}\partial^i\Phi'_1\partial_j\Phi'_1,
\end{align}
and 
\be
N^j_{\,\,\, i}=\frac{3}{2}\frac{1}{\nabla^2}
\left(\frac{\partial^j\partial_i}{\nabla^2}
-\frac{1}{3}\delta^j_{\,\,\, i}\right).
\ee
Eq.~\eqref{psi2} has to be solved with the superhorizon initial condition (with $\Phi_1({\bf x})\simeq -2\zeta_1({\bf x})/3$)~\cite{Inomata}
\be
\label{Psi2ic}
{\Psi_2}({\bf x})=
-2\Phi_1^2({\bf x})+2N^j_{\,\,\,i}\left(\partial^i\Phi_1({\bf x})\partial_j\Phi_1({\bf x})\right)=-\frac{8}{9}\zeta_1^2({\bf x})+\frac{8}{9}N^j_{\,\,\, i}\left(\partial^i\zeta_1({\bf x})\partial_j\zeta_1({\bf x})\right),
\ee
and in momentum space the solution is the sum of two pieces
\be
{\Psi_2}(\tau,{\bf k})={\Psi_2}({\bf k})T(k\tau)+\int\frac{{\rm d}^3p}{(2\pi)^3}u v I_\Psi(u,v,k\tau)\left(\frac{2}{3}\right)^2\zeta_1(\bf p)\zeta_1({\bf k}-{\bf p}),
\ee
where $u=|{\bf k}-{\bf p}|/k$, $v=p/k$, and the function $I_\Psi$
can be found in Refs.~\cite{Bartolo:2006cu,Bartolo:2006fj,Inomata}.
A detailed inspection of the time behaviour of $I_\Psi(u,v,k\tau)$ shows that ${\Psi_2}(\tau,{\bf k}) \propto a^{-2}$ due to the last term of Eq.~\eqref{psi2}~\cite{Inomata}.

Similarly, one can introduce the second-order curvature perturbation $\zeta_2$, which in the Newtonian gauge takes the form~\cite{Bartolo:2003bz,Malik:2008im,Inomata}
\begin{align}
	\label{eq:zeta_so_def}
	\zeta_2 =
	-\Psi_2 + \frac{1}{4} \delta_2 + \frac{1}{16 \mathcal H} \left(  \delta_1^2\right)' - \frac{1}{9} \delta_1^2   
	- \frac{1}{1 \mathcal H} \delta_1 \left( {\Phi'_1} + 2 \mathcal H \Phi_1 \right) - \frac{1}{32\mathcal H^2} \left[ \delta_1^{,k} \delta_{1 \, ,k} - \nabla^{-2} \left(  \delta_{1 \, ,i }\delta_{1 \, ,j} \right)^{,ij} \right].
\end{align}
On sub-horizon scales, where all gauges coincide and thus $\delta_1 = \delta_1^\text{\tiny com}$, this expression provides the behaviour for the curvature perturbation at second-order. The last term represents the leading contribution at late time and  one can show that 
\be
\frac{2}{3}\nabla^2\zeta_2(t,{\bf x})\simeq -\frac{1}{48a^2H^2}\sum_{i\neq j}\partial_i\partial_j\delta_1^\com(t,{\bf x})\partial^i\partial^j\delta_1^\com(t,{\bf x}),
\ee
as shown in Eq.~\eqref{nonspher} of the main text.

Finally, we provide the steps to estimate the second-order density contrast in the comoving gauge $\delta_2^\text{\tiny com}$, as shown in Eq.~\eqref{delta2com} of the manuscript, which is needed to compute the corresponding impact on the threshold of PBH formation. 
Starting from Eq.~\eqref{psi2} and treating the laplacian term $\nabla^2 \Psi_2$ as a perturbation, we can rewrite the equation by introducing a source term $\mathcal{S} (t, {\bf x}) $ as ${\ddot \Psi_2}+5 H{\dot \Psi_2} = \mathcal{S}$ (working with cosmic time), such that its solution formally reads
\begin{align}
\Psi_2(t,{\bf x})&=
\int^t\frac{{\rm d} t'}{a^5(t')}\int^{t'}{\rm d} t''a^5(t'')\frac{\mathcal{S} ({\bf x})}{a^2(t'')},
\end{align}
where the source has been evaluated with proper superhorizon initial conditions, since we are interested in evaluating the density profile at the horizon crossing time. The equation then simplifies to
\begin{align}
\Psi_2(t,{\bf x}) &= \Psi_2 ({\bf x})  + \int \d t \left( \frac{1}{15}\frac{\nabla^2\Psi_2}{a^2 H} + \frac{1}{a^2 H}\left[\frac{2}{15} \partial^i \Phi_1 \partial_i \Phi_1 + \frac{8}{15}\Phi_1\nabla^2\Phi_1
+ \frac{2}{5} N^j_{\,\,\, i} \nabla^2 \left(\partial^i\Phi_1\partial_j\Phi_1\right)\right] \right) \nonumber \\
&= \Psi_2 ({\bf x})  +  \frac{1}{30}\frac{\nabla^2\Psi_2 ({\bf x})}{a^2 H^2} + \frac{1}{a^2 H^2}\left[\frac{1}{15} \partial^i \Phi_1 \partial_i \Phi_1 + \frac{4}{15}\Phi_1 ({\bf x}) \nabla^2\Phi_1 ({\bf x})
+ \frac{1}{5} N^j_{\,\,\, i} \nabla^2 \left(\partial^i\Phi_1 ({\bf x}) \partial_j\Phi_1 ({\bf x})\right)\right].
\end{align}
By implementing the initial conditions shown in Eq.~\eqref{Psi2ic}, we get after some simplifications
\begin{align}
\label{solPsi2threshold}
\Psi_2(t,{\bf x})
& = -2\Phi^2_1({\bf x})+ 3 \frac{1}{\nabla^2} \left(\frac{\partial^j\partial_i}{\nabla^2}
-\frac{1}{3}\delta^j_{\,\,\, i}\right) \left(\partial^i\Phi_1({\bf x})\partial_j\Phi_1({\bf x})\right) -  \frac{1}{5}\frac{1}{a^2 H^2}  \partial^i \Phi_1 \partial_i \Phi_1   + \frac{1}{10}\frac{1}{a^2 H^2} \left(\frac{\partial^j\partial_i}{\nabla^2}\right)\left(\partial^i\Phi_1({\bf x})\partial_j\Phi_1({\bf x})\right) \nonumber \\
&  + \frac{2}{15} \frac{1}{a^2 H^2}\Phi_1\nabla^2\Phi_1  + \frac{3}{10} \frac{1}{\nabla^2}\frac{1}{a^2 H^2} \left(2\partial^j\nabla^2\Phi_1\partial_j\Phi_1 + \nabla^2\Phi_1\nabla^2\Phi_1 + \partial^i \partial^j\Phi_1 \partial_i \partial_j\Phi_1 
\right),
\end{align}
where we have used that $ \nabla^2 \Phi_1^2 = \partial^i \partial_i \Phi_1^2 = \partial^i (2 \Phi_1 \partial_i \Phi_1) = 2 \partial^i \Phi_1 \partial_i \Phi_1 + 2 \Phi_1 \nabla^2 \Phi_1$.
By taking a Laplacian, we get
\begin{align}
\nabla^2 \Psi_2(t,{\bf x}) &= -2\nabla^2 \Phi^2_1({\bf x}) + 3 \frac{\partial^j\partial_i}{\nabla^2} (\partial^i\Phi_1({\bf x})\partial_j\Phi_1({\bf x})) -  \partial^i \Phi_1({\bf x}) \partial_i \Phi_1({\bf x})\nonumber \\
 & - \frac{2}{5}\frac{1}{a^2 H^2} (\partial^i \nabla^2 \Phi_1 ({\bf x}) \partial_i \Phi_1 ({\bf x}) + \partial^i \partial_j \Phi_1 ({\bf x}) \partial_i \partial^j \Phi_1 ({\bf x})) \nonumber \\
 & + \frac{1}{10}\frac{1}{a^2 H^2} \left(2 \partial^j \nabla^2 \Phi_1({\bf x})\partial_j\Phi_1({\bf x}) + \nabla^2\Phi_1({\bf x})\nabla^2\Phi_1({\bf x}) + \partial^i \partial^j\Phi_1({\bf x}) \partial_i \partial_j\Phi_1({\bf x}) \right) \nonumber \\
 & + \frac{2}{15} \frac{1}{a^2 H^2}  \left( \nabla^2  \Phi_1 ({\bf x}) \nabla^2\Phi_1 ({\bf x}) + 2\partial_j  \Phi_1 ({\bf x}) \partial^j \nabla^2  \Phi_1 ({\bf x})  + \Phi_1 ({\bf x})\nabla^4 \Phi_1 ({\bf x}) \right) \nonumber \\
 & + \frac{3}{10} \frac{1}{a^2 H^2} \left(2\partial^j\nabla^2\Phi_1 ({\bf x})\partial_j\Phi_1 ({\bf x}) + \nabla^2\Phi_1 ({\bf x})\nabla^2\Phi_1 ({\bf x}) + \partial^i \partial^j\Phi_1 ({\bf x}) \partial_i \partial_j\Phi_1 ({\bf x}) \right),
\end{align}
such that
\begin{align}
& \nabla^2 \Psi_2(t,{\bf x}) = -2\nabla^2 \Phi^2_1({\bf x}) + 3 \frac{\partial^j\partial_i}{\nabla^2} (\partial^i\Phi_1({\bf x})\partial_j\Phi_1({\bf x})) -  \partial^i \Phi_1({\bf x}) \partial_i \Phi_1({\bf x})\nonumber \\
& + \frac{1}{a^2 H^2} \lp 
\frac{2}{3} \partial^i \nabla^2  \Phi_1({\bf x}) \partial_i \Phi_1({\bf x}) + \frac{8}{15} \nabla^2 \Phi_1({\bf x}) \nabla^2 \Phi_1({\bf x})   + \frac{2}{15} \Phi_1({\bf x}) \nabla^4  \Phi_1({\bf x})
\rp.
\end{align}
The second-order density contrast then becomes 
\begin{align}
\delta_2^\text{\tiny  com} &= \frac{2}{3} \frac{1}{a^2 H^2} \nabla^2 \Psi^{(2)}
 +2 \frac{1}{a^2 H^2} \partial_i \Phi^{(1)}\,\partial^i \Phi^{(1)}
+ \frac{16}{3} \frac{1}{a^2 H^2} \Phi^{(1)}\ \nabla^2 \Phi^{(1)} \nonumber \\
&  = \frac{1}{a^2 H^2} \left( -\frac{4}{3} \nabla^2 \Phi^2_1({\bf x}) + 2 \frac{\partial^j\partial_i}{\nabla^2} (\partial^i\Phi_1({\bf x})\partial_j\Phi_1({\bf x})) + \frac{4}{3} \partial^i \Phi_1({\bf x}) \partial_i \Phi_1({\bf x}) + \frac{16}{3} \Phi^{(1)}\ \nabla^2 \Phi^{(1)} \right) \nonumber \\
& + \frac{1}{a^4 H^4} \lp 
\frac{2}{3} \partial^i \nabla^2  \Phi_1({\bf x}) \partial_i \Phi_1({\bf x}) + \frac{8}{15} \nabla^2 \Phi_1({\bf x}) \nabla^2 \Phi_1({\bf x})   + \frac{2}{15} \Phi_1({\bf x}) \nabla^4  \Phi_1({\bf x})
\rp,
\end{align}
that provides Eq.~\eqref{delta2com} of the main text.

\end{document}